\documentclass[preprints,article,accept,pdftex,moreauthors,pdftex]{Definitions/mdpi} 
\firstpage{1} 
\pubvolume{1}
\issuenum{1}
\articlenumber{0}
\pubyear{2023}
\copyrightyear{2022}
\datereceived{} 
\daterevised{}
\dateaccepted{} 
\datepublished{} 
\hreflink{https://doi.org/} % If needed use \linebreak
\Title{On the determination of the 3D velocity field in terms of conserved variables in a compressible ocean}

\TitleCitation{On the determination of the 3D velocity field in terms of conserved variables in a compressible ocean}

\Author{R\'emi Tailleux $^{1}$\orcidA{}}

\AuthorNames{R\'emi Tailleux}

\AuthorCitation{Tailleux, R.}
\address{%
$^{1}$ \quad Department of Meteorology, University of Reading, Reading, UK}

\corres{Correspondence: R.G.J.Tailleux@reading.ac.uk}

\abstract{Explicit expressions of the 3D velocity field in terms of the conserved quantities of ideal fluid thermocline theory, namely Bernoulli function, density, and potential vorticity, are generalised here to a compressible ocean with a realistic nonlinear equation of state. The most general such expression is the `inactive wind' solution, an exact nonlinear solution of the inviscid compressible Navier-Stokes that satisfies the continuity equation as a consequence of Ertel's potential vorticity theorem. Such expressions are shown to be non-unique due to the non-uniqueness of the choice of Bernoulli function and in general approximately differ by the magnitude of their vertical velocity component. Due to the thermobaric nonlinearity of the equation of state, the expression of the 3D velocity field for a compressible ocean is found to resemble its ideal fluid counterpart only if constructed in terms of the available form of Bernoulli function as per Lorenz theory of available potential energy (APE). APE theory also naturally defines a quasi-material approximately neutral density variable called Lorenz reference density, which in turn defines a potential vorticity variable minimally affected by thermobaric production, thus providing all necessary tools for extending most results of ideal fluid thermocline theory to a compressible ocean. }


\begin{document}

%%%%%%%%%%%%%%%%%%%%%%%%%%%%%%%%%%%%%%%%%%
%\setcounter{section}{-1} %% Remove this when starting to work on the template.

\section{Introduction}

Although oceanic observations have increased dramatically over the past decades, especially since the development of the Argo floats program (https://argo.ucsd.edu), these still primarily constrain the density stratification rather than the velocity field. As a result, even in the most sophisticated ocean state estimate products \cite{Lee2009,Forget2015,Masuda2021,Wunsch2019}, the realism of the simulated 3D oceanic velocity field ${\bf v} = (u,v,w)$ can rarely if ever be ascertained directly. Indeed, even when current meters are available, these usually contain high-frequency motions linked to tides and gravity waves not usually captured by numerical ocean models thus complicating the comparison. In any case, comparisons when possible are only limited to the horizontal velocity field as the vertical velocity is generally too small to be directly measurable. 

\par 

Since for all practical purposes the 3D oceanic velocity field is not a measurable parameter of the system, the development of accurate dynamical theories predicting how to infer it from available oceanic observations plays an especially important role in physical oceanography. It is therefore no surprise that the subject has a long history going back to the early days of the discipline, which to date has primarily relied on the use of the dynamic method \cite{Fomin1964}. As is well known, the dynamic method assumes geostrophic and hydrostatic balances and provides an explicit expression of the horizontal velocity field in terms of vertical integrals of the horizontal density gradients relative to the velocity field at some reference level. Understanding how to specify the latter is one of the main challenges of the dynamic method and has given rise to many different approaches to tackle it. The simplest approach and most often encountered one in the literature assumes the existence of a level of no motion at which the geostrophic velocity vanishes, see \cite{OlbersWillebrand1984} for a discussion of its limitations. Since then, such an approach has been superseded by the beta-spiral method of \cite{Schott1978} or the inverse method(s) of \cite{Wunsch1977,Wunsch1978}, whose connection is discussed in \cite{Davis1978}. \cite{Behringer1979,Behringer1980,Killworth1986,Bigg1985} are examples of studies further discussing these ideas. At some point, there was hope that the sea surface height (SSH) signal measured by satellite altimetry would be accurate enough to constrain the surface geostrophic velocity field. However, this requires a more accurate determination of the geoid than is presently available. Moreover, the fact that the SSH signal is contaminated by small-scale and high-frequency processes, while also reflecting transient variations of the interior density field, complicates filtering out only the relevant part needed by the dynamic method \cite{Park2004}. More recently, measurements of Argo floats displacements have shown promise for constraining the geostrophic velocity at the floats' parking depth, as recently demonstrated by \cite{acdv2016,acdv2019}. 
Nevertheless, some limitations remain as Argo floats displacements are also a priori impacted by energetic small scale and transient ageostrophic motions in addition to the geostrophic flow, which the authors sought to mitigate by averaging over the Argo period. Outside the Argo period, or when focusing on time snapshots, alternative/complementary approaches for constraining the unknown reference level are still needed. 

\par 

While the geostrophic approximation is generally accepted to represent an accurate theory for the horizontal velocity field ${\bf u} = (u,v)$, how best to predict the vertical velocity field $w$ is in contrast much less understood and still actively debated. Physically, there are two main fundamental approaches for thinking about the vertical velocity field. The first approach, which as far as we are aware underlies the computation of the vertical velocity field in all existing numerical ocean models \cite{Griffies2004}, is based on vertically integrating the continuity equation $\nabla_h \cdot {\bf u} + w_z = 0$. Doing so yields an expression for $w$ controlled by the horizontal velocity divergence relative to the vertical velocity at some reference level $z_r$ 
\begin{equation}
       w (x,y,z,t) = \underbrace{w(x,y,z_r,t)}_{w_r} - \int_{z_r}^z 
       \nabla_h \cdot {\bf u} \,{\rm d}z' ,
       \label{w_from_continuity} 
\end{equation}
where $\nabla_h$ is the horizontal nabla operator. As is well known, using the continuity equation in conjunction with geostrophy yields the celebrated linear Sverdrup balance $\beta v = f \partial w/\partial z$, which may be integrated to yield
\begin{equation}
      w = w_r + \frac{\beta}{f} \int_{z_r}^z v \,{\rm d}z' ,  
      \label{linear_sverdrup_balance}
\end{equation}
see \cite{Wunsch2011}, where $f$ is the Coriolis parameter, and $\beta = df/dy$, with $y$ denoting latitude. The second approach is based on extracting $w$ from the conservation equation $\partial C/\partial t + {\bf v}\cdot \nabla C = \dot{C}$ of any conserved tracer $C$, viz., 
\begin{equation}
       w = \left ( \frac{\partial C}{\partial z} \right )^{-1} 
       \left ( \dot{C} - \frac{\partial C}{\partial t} 
       - {\bf u} \cdot \nabla_h C \right ) ,
       \label{w_from_tracer} 
\end{equation}
where $\dot{C}$ represents the non-material sinks/sources of $C$. Both approaches come with important theoretical challenges. In the first approach, the difficulty stems from the fact that the horizontal divergence $\nabla\cdot {\bf u}$ is often dominated by the ageostrophic component ${\bf u}_{ag}$ of the velocity, in which case (\ref{linear_sverdrup_balance}) may become very inaccurate. In the second approach, the difficult stems from that the time derivative $\partial C/\partial t$ and (perhaps to a lesser extent) the diabatic term $\dot{C}$ may both be important for predicting the large-scale vertical velocity field, yet often hard (if at all possible) to constrain observationally. Existing approaches, therefore, can be regarded as attempts to mitigate such difficulties in some way. For instance, the well-known omega equation is obtained by artfully combining two prognostic equations (for vorticity and buoyancy respectively) in a way that eliminate the time-derivative in each equation
to formulate a diagnostic elliptic problem for the vertical velocity, see \cite{BuongiornoNardelli2020} for a recent implementation. Another approach of interest showing how to infer the vertical velocity from individual moorings is that proposed by \cite{Sevellec2015}. 

\par 
Physically, one key reason that makes the second approach potentially the most attractive one is because if one ignores the difficulties associated with the time-dependent and diabatic terms $\partial C/\partial t$ and $\dot{C}$ in (\ref{w_from_tracer}), knowledge of the geostrophic velocity field is generally sufficient enough for accurately estimating the horizontal advection term $-{\bf u}\cdot \nabla C$. In contrast, knowledge of the ageostrophic velocity component ${\bf u}_{ag}$ is often needed for an accurate determination of the horizontal velocity divergence in the first approach, as mentioned above. As a result, there has been much interest in seeking 
to exploit the existence of quasi-material conserved quantities to construct explicit expressions of the steady-state 3D velocity field. Thus, if $C_1$ and $C_2$ represent two independent conserved quantities, then it is well known that in a steady state the 3D velocity should lie at the intersection of two iso-surfaces of $C_1$ and $C_2$. Mathematically, this implies that the velocity field may be written in the form 

\begin{equation}
      {\bf v} = \lambda \nabla C_1 \times \nabla C_2 ,
      \label{v_c1_c2} 
\end{equation}

\noindent 
for some scalar field $\lambda$. 
Physically, the constraint determining $\lambda$ is that the horizontal component of (\ref{v_c1_c2}) be geostrophic either exactly or approximately. The conserved quantities $C_1$ and $C_2$ then determine the vertical velocity according to $w = {\bf u} \cdot {\bf S}_1$ or $w = {\bf u}\cdot {\bf S}_2$ or both, where ${\bf S}_1$ and ${\bf S}_2$ are the horizontal `slope' vectors
\begin{equation}
       {\bf S}_i  = - \left ( \frac{\partial C_i}{\partial t}
       \right )^{-1} {\bf u} \cdot \nabla_h C_i, \qquad
       i = \{1,2 \} .
\end{equation}
Thus, depending on the conserved quantities considered, the relations $w = {\bf u}\cdot {\bf S}_1$ and $w = {\bf u}\cdot {\bf S}_2$ may or may not define the same vertical velocity. In practice, this means that the problem to be solved might be over-determined and that its resolution may need the use of least-squares. 

\par 
While (\ref{v_c1_c2}) can in principle be implemented in practice using a variety of conserved tracers, three particular quantities deserve special attention owing to their theoretical and dynamical importance in oceanography, namely: the Bernoulli function $B$, potential vorticity $Q$, and density $\rho$. Physically, this is because these quantities are the ones underlying the ideal fluid thermocline equtions \cite{Welander1971b} that have formed the basis for most ocean circulation theories, as well as those originally considered by Needler \cite{Needler1985}, who pioneered the use of (\ref{v_c1_c2}). Unfortunately, such quantities have been so far unambiguously defined only for an ideal fluid, which significantly hinder our ability to evaluate the usefulness of (\ref{v_c1_c2}) for the real compressible ocean. Although several studies have extended Needler's approach since, e.g., 
\cite{Chu1995,Chu2000,Kurgansky2002,Kurgansky2021}, most discussions with a few exceptions, e.g., \cite{McDougall1995,Ochoa2020} have remained limited to the case of an ideal fluid in geostrophic balance. To make progress, this paper aims to show how to generalise Needler's approach to the case of a fully compressible ocean, in order to be able to test its usefulness systematically in the future (which is beyond the scope of this paper). Section \ref{ideal_fluid} reviews the basic properties of the ideal fluid thermocline equations and introduce the concept of 'available' Bernoulli function. Section \ref{compressible_fluid} shows how to generalise ideal fluid thermocline theory to compressible seawater, and shows that the results is related to \cite{Gassmann2014} `inactive wind' solution. Section \ref{discussion} discusses the results and some perspectives. 

\section{Absolute velocity field based on ideal fluid thermocline theory} 
\label{ideal_fluid}

\subsection{Thermodynamic form of ideal fluid thermocline equations}

We begin by briefly reviewing the properties and structure of the ideal fluid thermocline equations \cite{Welander1971b} that underlie Needler's determination of the absolute velocity field in terms of conserved properties \cite{Needler1985}. Although most of the material is standard, an important new element is the introduction of the concept of `available' Bernoulli function, where the term `available' is meant to parallel that of Lorenz `available potential energy' \cite{Lorenz1955}. The ideal fluid thermocline equations are here written in the following form

\begin{equation}
          f {\bf k} \times \rho {\bf v} + \nabla p = -\rho g {\bf k} ,
          \label{ife_momentum} 
\end{equation}

\begin{equation} 
       \nabla \cdot {\bf v} = 0 ,
       \label{ife_continuity}
\end{equation}

\begin{equation} 
     {\bf v}\cdot \nabla \rho = 0 ,
     \label{rho_conservation} 
\end{equation} 

\noindent 
and respectively describe the geostrophic and hydrostatic momentum balance, continuity, and conservation of density respectively, where ${\bf v} = (u,v,w)$ is the three-dimensional velocity field, $p$ is pressure, $\rho$ is density, $g$ is the acceleration of gravity, $f$ is the Coriolis parameter, and ${\bf k}$ is the unit vector pointing upward. While the geostrophic and hydrostatic balances are expected to meaningfully describe the balanced part of the flow even under transient evolution for sufficiently small Rossby number, the steady form of the density equation (\ref{rho_conservation}) is in contrast much more difficult to justify rigorously, especially without clarity as to what density variable $\rho$ is supposed to represent. Presumably, any justification of (\ref{rho_conservation}) must require consideration of some form of temporal averaging that does not introduce eddy-correlation terms, such as thickness-weighted averaging \cite{Young2012}. As the issue is quite complex and cannot fully be addressed without having first clarified the exact nature of $\rho$, its full treatment is deferred to a subsequent study.

The main key step for linking the velocity field to the conserved quantities of the system is to rewrite the momentum 
(\ref{ife_momentum}) in its thermodynamic form (also known as the Crocco-Vazsonyi form \citep{Crocco1937,Vazsonyi1945}), 

\begin{equation}
         f {\bf k} \times \rho {\bf v} + \rho_{\star} 
         \nabla B_h^{ideal} = \rho_{\star} {\bf P}_h^{ideal}  , 
         \qquad {\bf P}_h^{ideal}  = \frac{g z \nabla \rho}{\rho_{\star}}  ,
         \label{EGTBF_boussinesq} 
\end{equation}

\noindent 
where $B_h^{ideal} = (p + \rho g z)/\rho_{\star}$ 
is the standard Bernoulli function and $\rho_{\star}$ is a constant Boussinesq reference density.

\subsection{Available and background Bernoulli functions}
A key idea of this paper is that Lorenz APE theory holds the key to understanding how to generalise the key ingredients of ideal fluid thermocline theory, that is the Bernoulli function, potential vorticity, and density, to a compressible ocean. The key ingredients needed here are the reference pressure and density profiles $p_0(z)$ and $\rho_0(z)$ characterising Lorenz reference state of minimun potential energy obtainable by means of an adiabatic an isohaline re-arrangement of mass, and the reference position of a fluid parcel $z_r$ defined as a solution of the level of neutral buoyancy (LNB) equation $\rho = \rho_0(z_r)$, which defines $z_r = z_r(\rho)$ as a function of density only \cite{Tailleux2013b,Saenz2015,Tailleux2018}. Physically, APE theory is useful for providing a physical mean to tease apart the part of the potential energy that is available for reversible conversions with kinetic energy (the APE) from the dynamically inert part (the BPE) that is not. Because the Bernoulli function $B_h^{ideal} = (p + \rho g z)/\rho_{\star}$ is thermodynamic in nature, the same idea that only a fraction of it is available for reversible conversions with kinetic energy must apply. This motivates us to define its dynamically inert part in a Lagrangian sense as its value in Lorenz reference state, viz., 

\begin{equation}
     B_r^{ideal} = \frac{p_0(z_r) + \rho_0(z_r) g z_r}{\rho_{\star}}  ,
     \label{inert_bernoulli} 
\end{equation}

\noindent 
and the `available' part of the Bernoulli function as 

\begin{equation}
\begin{split} 
      \rho_{\star} B_a^{ideal} = \rho_{\star}( B_h^{ideal} - B_r^{ideal} )
      = &  p-p_0(z) + p_0(z)-p_0(z_r) 
      + \rho g (z-z_r) \\ 
      = &  p-p_0(z) + \underbrace{\int_{z_r}^z 
      g (\rho-\rho_0(z'))\,{\rm d}z'}_{\rho_{\star} E_a} 
      \label{ba_ideal_definition} 
\end{split} 
\end{equation}

\noindent 
using the fact that $p_0'(z)=-g \rho_0(z)$ and $\rho_0(z_r)=\rho$ by definition, where the underbraced quantity $E_a$ can be recognised as the positive definite APE density originally introduced by \cite{Holliday1981} and \cite{Andrews1981} and later extended to multi-component Boussinesq and stratified fluids by \cite{Tailleux2013b,Tailleux2018}. Physically, the APE density may be written as the work against buoyancy forces needed to bring a fluid parcel from its level of neutral buoyancy $z_r$ (hence satisfying $b(S,\theta,z_r)=0))$ to its actual position, viz.,  
\begin{equation}
         E_a = - \int_{z_r}^z b(S,\theta,z')\,{\rm d}z' ,
\end{equation}
the only difference between the expressions for a Boussinesq and general compressible fluid being in the expressions for the buoyancy $b$. For a Boussinesq fluid, $b = -g(\rho-\rho_0(z))/\rho_{\star}$, while for compressible seawater discussed in next section $b=-g[1-\rho_0(z)\upsilon(S,\theta,p_0(z))]$. Note that since $\rho = \rho_0(z_r)$, $\rho_{\star} E_a$ may also be rewritten as 

\begin{equation}
     \rho_{\star} E_a = - g \int_{z_r}^z \int_{z_r}^{z'} 
     \frac{d\rho_0}{dz} (z'') \,{\rm d}z'' \,{\rm d}z' ,
\end{equation}

\noindent so that for small departure from the reference position,

\begin{equation}
    E_a  \approx - \frac{g}{\rho_{\star}} \frac{d\rho_0}{dz}(z_r) \frac{(z-z_r)^2}{2} = \frac{N_r^2 (z-z_r)^2}{2} 
\end{equation}

\noindent 
which the reader unfamiliar with APE may still recognise, and which makes the positive definite character of $E_a$ clearer if needed. 

\par 
 
If we remove $\nabla B_r$ from both sides of (\ref{EGTBF_boussinesq}), using the easily verified result that $\nabla B_r = g z_r \nabla \rho$, yields the following available thermodynamic form of momentum balance 

\begin{equation}
      f {\bf k} \times \rho {\bf v} + \rho_{\star} \nabla B_a^{ideal}
      = \rho_{\star} {\bf P}_a^{ideal},
      \qquad \rho_{\star} {\bf P}_a^{ideal} = g (z-z_r) \nabla \rho .
      \label{EGTBF_boussinesq_available} 
\end{equation} 

\noindent 
A key point to note here is that in both (\ref{EGTBF_boussinesq}) and (\ref{EGTBF_boussinesq_available}), the two P-vectors ${\bf P}_h^{ideal} $ and ${\bf P}_a^{ideal}$ are proportional to the gradient of density $\nabla \rho$, and therefore perpendicular to the isopycnal surfaces $\rho = {\rm constant}$. We shall see in Section \ref{compressible_fluid} that this property is lost in a compressible ocean. It is of interest to note that ${\bf P}_h^{ideal} $ is the ideal fluid counterpart of the P-vector previously identified by \cite{Nycander2011}.

\subsection{Bernoulli and Potential vorticity (PV) theorems} 

The Bernoulli theorem \cite{Schar1993} and Ertel's PV conservation theorem \cite{Muller1995,Schubert2004} play a key role in this paper. As regards to the former, it is trivially obtained by taking the inner product of (\ref{EGTBF_boussinesq}) and (\ref{EGTBF_boussinesq_available}) by ${\bf v}$, accounting for (\ref{rho_conservation}), which immediately yields ${\bf v}\cdot \nabla B_h = 0$ and ${\bf v}\cdot \nabla B_a = 0$. The proof of the potential vorticity (PV) conservation theorem is somewhat more involved. To obtain it, first divide (\ref{ife_momentum}) by $\rho$ and take the curl thus leading to

\begin{equation}
      {\bf k} ( {\bf v} \cdot \nabla f ) - 
      f \frac{\partial{\bf v}}{\partial z} = 
       \frac{\nabla \rho \times \nabla p}{\rho^2} .
       \label{intermediate} 
\end{equation}

\noindent 
Next, take the inner product of (\ref{intermediate}) by $\nabla \rho$, which then yields

\begin{equation}
      \frac{\partial \rho}{\partial z} {\bf v} \cdot \nabla f
      - f \nabla \rho \cdot \frac{\partial {\bf v}}{\partial z} = 
      {\bf v} \cdot \nabla \left ( f \frac{\partial \rho}{\partial z} \right ) = 0 ,
\end{equation}

\noindent 
which establishes the material conservation of the potential vorticity (PV) $Q= (f/\rho_{\star}) \partial \rho/\partial z$ as expected.

\subsection{Determination of the absolute velocity field in terms of conserved quantities} 

Taking the cross product of $\nabla \rho$ with (\ref{EGTBF_boussinesq}) or (\ref{EGTBF_boussinesq_available}), and using the result that
$\nabla \rho \times ( f {\bf k} \times {\bf v} ) = 
({\bf v}\cdot \nabla \rho )  f {\bf k} - (f \nabla \rho \cdot {\bf k} ) {\bf v} = -f\rho_z {\bf v}$ (which follows from the vector algebra relation
${\bf A} \times ({\bf B} \times {\bf C}) = ({\bf A}\cdot {\bf C}) {\bf B} - ({\bf A}\cdot {\bf B}){\bf C}$ and conservation of density ${\bf v}\cdot \nabla \rho=0$, with $\rho_z$ being shorhand for $\partial \rho/\partial z$) leads after some manipulation to the following explicit expression of ${\bf v}$ previously obtained by \cite{Needler1985},  

\begin{equation}
     \rho {\bf v} = \frac{\nabla \rho \times \nabla B}{Q}, 
     \label{needler_formula} 
\end{equation}

\noindent 
which explicitly relies on density and the Bernoulli function being conserved following fluid parcels, regardless of which form of Bernoulli function is used ($B$ refers indifferently to $B_h$ or $B_a$). 

\par 
To show that (\ref{needler_formula}) naturally satisfies the continuity equation as a consequence of Ertel's PV conservation theorem, simply take its divergence, which yields

\begin{equation}
     \nabla \cdot (\rho {\bf v}) = -\frac{(\nabla \rho \times \nabla B)\cdot \nabla Q}{Q^2} = - \frac{\rho {\bf v}\cdot \nabla Q}{Q} = 0 ,
     \label{proof_continuity}
\end{equation}

\noindent 
QED. As a result, (\ref{needler_formula}) represents an exact steady solution of the ideal fluid equations (\ref{ife_momentum}-\ref{rho_conservation}) that also predicts the vertical velocity

\begin{equation}
        \rho w = \frac{{\bf k}\cdot (\nabla_h \rho \times \nabla_h B)}{Q} ,
        \label{w_needler} 
\end{equation}

\noindent 
not just the horizontal velocity, where $\nabla_h$ denotes the horizontal gradient. Importantly, Needler's formula (\ref{needler_formula_2}) is insensitive to the particular choice of Bernoulli function --- $B$ or $B_a$  --- used to estimate it due to $B_r = B_r(\rho)$ being a function of $\rho$ only.

\subsection{Bernoulli method} 

Physically, one key reason why Needler's formula (\ref{needler_formula}) is appealing is because the idea that the steady or time-averaged 3D velocity field should lie at the intersection of the iso-surfaces of two conserved quantities is a priori valid beyond the geostrophic approximation. Its other key advantage is that it naturally satisfies the continuity equation as a consequence of Ertel potential vorticity conservation theorem, which also generalises beyond the geostrophic approximation, as discussed in next section. In the context of the dynamic method, however, Needler's formula does not in itself solve the problem of the unknown reference level. Indeed, because $\nabla_h  B = \nabla p_h + g z \nabla_h \rho$, it is easily verified that the horizontal velocity that it predicts is the standard geostrophic balance, while the vertical velocity predicted by (\ref{w_needler}) is actually the one predicted more easily from the density equation ${\bf v}\cdot \nabla \rho = 0$, i.e.,

\begin{equation}
      w = -\left ( \frac{\partial \rho}{\partial z} \right )^{-1} 
      {\bf u} \cdot \nabla_h \rho = {\bf u}\cdot {\bf S}_{\rho} , 
\end{equation}

\noindent 
where ${\bf S}_{\rho}$ is the slope vector associated with the isopycnal surfaces. In other words, Needler's formula does not solve the problem of the unknown reference level because the Bernoulli function depends on the same unknown constant of integration as the pressure field.

\par 
To circumvent the difficulty and
turn (\ref{needler_formula}) into something concretely useful, it is necessary to invoke the result that if $\rho$, $B$, and $Q$ are all conserved along fluid parcel trajectories, they must be functionally related. Thus, to the extent that such a result holds and that this functional relationship may be written in the form $B = G(\rho,Q)$, it become possible to rewrite (\ref{needler_formula}) as 

\begin{equation}
       \rho {\bf v} = \frac{\partial G}{\partial Q}
       \frac{\nabla \rho \times \nabla Q}{Q} ,
       \label{needler_formula_2} 
\end{equation}

\noindent 
\cite{Needler1985}. Now, the key advantage of 
(\ref{needler_formula_2}) over (\ref{needler_formula}) is that the term $(\nabla \rho \times \nabla Q)/Q$ can in principle be empirically estimated from climatological fields without having to solve the problem of the unknown reference level, since the latter neither affects $\rho$ nor $Q$. In other words, (\ref{needler_formula_2}) provides a determination of the absolute velocity field up to the multiplicative constant $\partial G/\partial Q$. In this approach, the problem of the unknown reference level therefore transforms into the problem of how best to evaluate the functional relationship $G(\rho,Q)$ and the partial derivative $\partial G/\partial Q$. For related discussions of these ideas, the reader is referred to, e.g., \cite{Killworth1979,Killworth1980,Killworth1986,Kurgansky2002,Kurgansky2021,Chu1995,Chu2000,McDougall1995}. However, it seems fair to say that the current empirical evidence for a well defined relationship $B=B(\rho,Q)$ is inconclusive at best. Current approaches, however, have relied on using the conventional form of Bernoulli function $B=B_h$ rather than its available form $B=B_a$. It will be of interest in future studies to test whether using $B_a$, as well as the density variable discussed in next section, can lead to a better defined relationship $B_a = B_a (\rho,Q)$. 

\section{Generalisation to compressible seawater} 
\label{compressible_fluid} 

\subsection{Governing equations for compressible seawater} 

We now turn to a realistic nonlinear ocean described by the compressible Navier-Stokes equations, treating seawater as a two-constituent stratified fluid: 

\begin{equation}
      \frac{D{\bf v}}{Dt} + \boldsymbol{2 \Omega} \times {\bf v} 
      + \frac{1}{\rho}\nabla p = - \nabla \Phi + {\bf F} ,
      \label{true_momentum_balance} 
\end{equation} 

\begin{equation}
        \frac{\partial \rho}{\partial t} + \nabla \cdot (\rho {\bf v}) = 0 ,
\end{equation}

\begin{equation}
       \frac{D\eta}{Dt} = \dot{\eta}, \qquad \frac{DS}{Dt} = \dot{S} ,
\end{equation} 

\begin{equation}
   \upsilon = \frac{1}{\rho} = \upsilon(\eta,S,p) , 
\end{equation} 

\noindent 
where $\eta$ is the specific entropy, $S$ is salinity, $\upsilon = 1/\rho$ is the specific volume, ${\bf F}$ is a friction force, and $\Phi = g z$ is the geopotential. Moreover, $\dot{\eta}$ and $\dot{S}$ denote the diabatic sources/sinks of heat and salt due to molecular diffusive fluxes (radiation can be included if needed). As in the previous section, the first step is to rewrite the momentum balance (\ref{true_momentum_balance}) in its thermodynamic or Crocco-Vazsonyi form

\begin{equation}
     \frac{\partial {\bf v}}{\partial t} 
     + \boldsymbol{\omega}_a \times {\bf v} + \nabla B_h = {\bf P}_h + {\bf F},
\label{crocco_NSE} 
\end{equation}

\noindent 
where $\boldsymbol{\omega}_a = \boldsymbol{\xi} + 2 \boldsymbol{\Omega}$ is the absolute vorticity, $\boldsymbol{\xi} = \nabla \times {\bf v}$ is the relative vorticity. The result was obtained by using the well known relation $({\bf v}\cdot \nabla ){\bf v} = \nabla ({\bf v}^2/2) + \boldsymbol{\xi} \times {\bf v}$ as well as the total differential for specific enthalpy ${\rm d} h = T {\rm d}\eta + \mu {\rm d}S + \upsilon {\rm d}p$, where $T$ is the in-situ temperature, $\mu$ the relative chemical potential. For a compressible ocean, the conventional form of Bernoulli function is 

\begin{equation}
     B_h = \frac{{\bf v}^2}{2} + h(\eta,S,p) + \Phi(z) ,
\end{equation} 

%\begin{equation}
%\begin{split} 
%       B_h = &  \frac{{\bf v}^2}{2} + h(\eta,S,p) + \Phi(z) \\
%          = & \frac{{\bf v}^2}{2} + \Pi_1 + h(\eta,S,p_0(z)) + \Phi(z) 
%          + \frac{p-p_0(z)}{\rho} 
%\end{split} 
%\end{equation}

\noindent 
while the P-vector ${\bf P}_h$ is the vector that retains the thermohaline gradient of $h$, viz., 

\begin{equation}
       {\bf P}_h = \frac{\partial h}{\partial \eta} \nabla \eta  
       + \frac{\partial h}{\partial S} \nabla S = T \nabla \eta  
       + \mu \nabla S .
       \label{conventional_p_vector} 
\end{equation} 

\noindent 
As in the previous section, we introduce the available Bernoulli function as $B_a = B_h - B_r$, that is, as the difference between the conventional form of Bernoulli function $B_h$ and the background Bernoulli function

\begin{equation}
     B_r = h(\eta,S,p_0(z_r)) + \Phi(z_r) 
\end{equation}

\noindent 
where as before, $z_r = z_r(\eta,S)$ is the reference height of a fluid parcel in Lorenz reference state of minimum potential energy, which in practice may be computed using the computationally efficient algorithm of \citet{Saenz2015}. Physically, $z_r$ is as before defined as a solution of the level of neutral buoyancy equation, which for two-component compressible seawater takes the form 

\begin{equation}
      \upsilon(\eta,S,p_0(z_r)) = \upsilon_0(z_r) ,
      \label{lnd_equation_compressible} 
\end{equation}

\noindent 
where $\upsilon_0(z) = 1/\rho_0(z)$. The available Bernoulli function may thus be written in the form

\begin{equation}
\begin{split} 
      B_a = &  \frac{{\bf v}^2}{2} + h(\eta,S,p) - h(\eta,S,p_r) + g(z-z_r) 
      \\ 
       = & \frac{{\bf v}^2}{2} + \Pi + 
       \frac{p-p_0(z)}{\rho} 
\end{split} 
\end{equation}

\noindent 
where $\Pi = \Pi_1+\Pi_2$ is the potential energy density defined in \cite{Tailleux2018}, with $\Pi_1$ and $\Pi_2$ the subcomponents

\begin{equation}
    \Pi_1 = h(\eta,S,p)-h(\eta,S,p_0(z)) + \frac{p_0(z)-p}{\rho} ,
\end{equation}

\begin{equation} 
    \Pi_2 = h(\eta,S,p_0(z)) - h(\eta,S,p_0(z_r)) + 
    g(z-z_r)
\end{equation} 

\noindent 
As explained and demonstrated in \cite{Tailleux2018}, both $\Pi_1$ and $\Pi_2$ are positive definite. $\Pi_1$ may be interpreted as the available compressible energy (ACE), which represents the expansion/contraction work required to compress/expand from the reference pressure $p_0(z)$ to the actual pressure. As to $\Pi_2$, it is the APE density and represents the work against buoyancy forces required to move the fluid parcel from its reference position $z_r$ at pressure $p_0(z_r) = p_r$ to its actual position $z$ at pressure $p_0(z)$. As before, the available thermodynamic form of momentum balance is obtained by removing $\nabla B_r = T_r \nabla \eta + \mu_r \nabla S$ on both sides of (\ref{crocco_NSE}), which leads to

\begin{equation}
        \frac{\partial {\bf v}}{\partial t} 
        + \boldsymbol{\omega}_a \times {\bf v} + \nabla B_a 
        = {\bf P}_a + {\bf F} 
        \label{crocco_NSE_bis} 
\end{equation}

\noindent 
where the P-vector ${\bf P}_a$ takes the form

\begin{equation}
       {\bf P}_a = (T-T_r) \nabla \eta + (\mu-\mu_r) \nabla S .
       \label{available_p_vector} 
\end{equation} 
Note that the suffix 'r' denote thermodynamic quantities estimated at the reference pressure $p_r = p_0(z_r)$. 

\subsection{Comparison of ideal and compressible forms of Bernoulli functions and P-vectors} 

In an ideal fluid, Needler's formula (\ref{needler_formula}) defines the 3D velocity field as being perpendicular to the density surfaces $\rho={\rm constant}$ regardless of which particular form of $(B,{\bf P})$ is used, due to ${\bf P}_h^{ideal}$ and ${\bf P}_a^{ideal}$ being both parallel to $\nabla \rho$ in that case. The 3D velocity field is also perpendicular to the iso-surfaces of the Bernoulli function in both cases, but the iso-surfaces of $B_h^{ideal}$ and $B_a^{ideal}$ should in general appear distinct from each other. Whether the available form $(B_a^{ideal},{\bf P}_a^{ideal})$ is superior over the conventional form $(B_h^{ideal},{\bf P}_h^{ideal})$ cannot be determined from theoretical considerations alone; however, it is possible that the assumed empirical functional relationship $B=B(\rho,Q)$ underlying the Bernoulli method might be more accurately achieved in climatological observations of the density field for one of $(B,{\bf P})$ forms. 

\par 

The situation is different for compressible seawater, however, as ${\bf P}_a$ and ${\bf P}_h$ now define different directions in general, as is clear from their expressions (\ref{conventional_p_vector}) and (\ref{available_p_vector}). Moreover, it is easily verified that both P-vectors have a non zero helicity $H_p = {\bf P}\cdot (\nabla \times {\bf P})$ because of thermobaricity and are therefore both non-integrable. Mathematically, this means that it is not possible to identify a well defined seawater variable whose iso-surfaces are everywhere perpendicular to ${\bf P}$. From a theoretical viewpoint, this is both interesting and important, as it suggests that the superior form of $(B,{\bf P})$ may be the one for which the compressible and ideal expressions resemble the most to each other. 

\par 

\begin{figure}
    \centering
    \includegraphics[width=14cm]{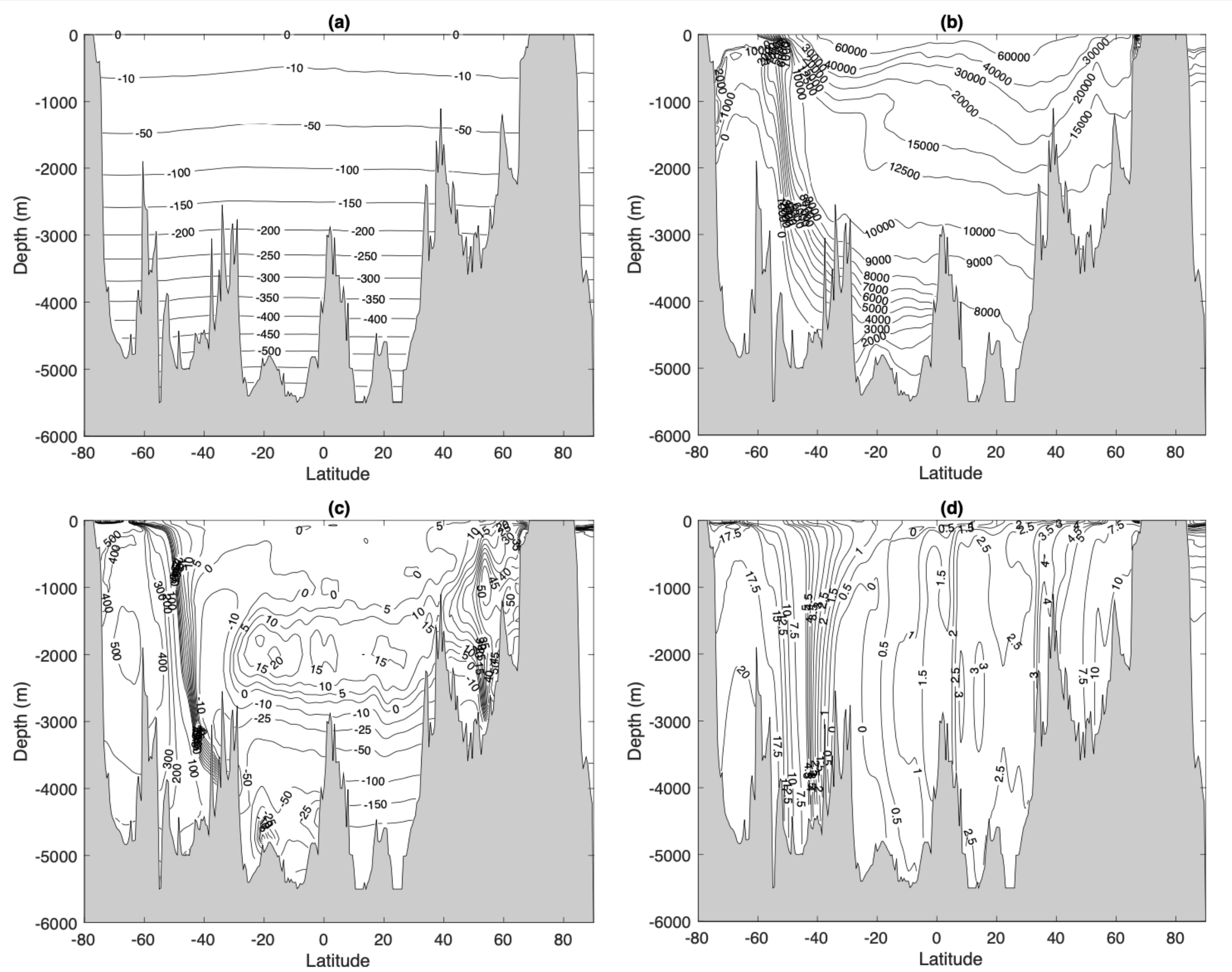}
    \caption{Illustrative examples of the different kinds of Bernoulli functions considered in this paper (in ${\rm J.kg^{-1}}$) along the $30^{\circ}W$ meridonal section in the Atlantic Ocean, with the hydrostatic pressure assuming a level of no motion at $1500\,{\rm m}$.  
    (a) $B_h^{ideal}$; (b) $B_h$; (c) $B_a^{ideal}$; (d) $B_a$.} 
    \label{fig:bernoulli}
\end{figure}

To explore this idea, table \ref{tab1} summarises the forms of Bernoulli function and P-vectors for an ideal and compressible forms established in the previous sections, while Fig. \ref{fig:bernoulli} illustrates the 4 different types of Bernoulli functions along the $30^{\circ}W$ meridional section in the Atlantic Ocean. We used the WOCE climatological dataset \cite{Gouretski2004} and computed the hydrostatic pressure assuming a level of no motion at $1500\,{\rm m}$. The Lorenz reference density and pressure profiles, as well as the computation of the reference depths, were computed as in \cite{Tailleux2021}. The contribution from the kinetic energy was ignored. This figure clearly shows that the Bernoulli function depends sensitively on the approach considered as well as on the choice of the arbitrary constants entering the definition of the specific enthalpy as defined by TEOS-10 (www.teos-10.org). Panel (a) shows the ideal standard Bernoulli function to be dominated by its depth variations, with values increasing with height from $-700\,{\rm J.kg^{-1}}$ at depth to close to $0$ at the surface. In contrast, the compressible standard Bernoulli function (Panel (b)), which depends on the TEOS-10 definition of specific enthalpy, is dominated by thermal variations $\propto c_{p0} \theta$ near the surface that results in much bigger values overall, which also increase with height from about close to zero at depth to about $60,000 \,{\rm J.kg^{-1}}$ near the surface. The ideal and compressible forms of Bernoulli function are depicted in Panels (c) and (d) respectively. Although our theory predicts that the ideal and compressible available Bernoulli functions should be close to each other, this is not the case in practice because our prediction (\ref{ba_ideal_definition}) assumes an incompressible $\rho$, which is not true of in-situ density. Panel (d) shows that the compressible available Bernoulli function $B_a$ is dominated by horizontal variations, which seems an optimal behaviour for plotting it on isopycnal surfaces as will be discussed in details in a subsequent study. Note that the range of values exhibited by $B_a$, comprised between 0 and $20\,{\rm J.kg^{-1}}$ approximately, is considerably smaller than that of all other Bernoulli functions, confirming the idea that $B_a$ is the only one not affected by irrelevant dynamical information, which we hope can be exploited in the future.

\begin{table}[H] 
\caption{Comparison of the ideal and compressible forms of Bernoulli function and P-vectors. Note that the Bernoulli functions for a compressible fluid don't include the kinetic energy term. \label{tab1}}
\newcolumntype{C}{>{\centering\arraybackslash}X}
\begin{tabularx}{\textwidth}{CCC}
\toprule
\textbf{Quantity}	& \textbf{Ideal}	& \textbf{Compressible}\\
\midrule
$B_h$	& $\frac{p+\rho gz}{\rho_{\star}}$	
& $h(\eta,S,p) + g z $\\
$B_a$		& $E_a + \frac{p-p_0(z)}{\rho_{\star}}$ 
& $\Pi_1 + \Pi_2 + \frac{p-p_0(z)}{\rho}$ \\
${\bf P}_h$ &  $\frac{gz\nabla \rho}{\rho_{\star}}$ & 
     $T \nabla \eta + \mu \nabla S$ \\
${\bf P}_a$ & $\frac{g(z-z_r) \nabla \rho}{\rho_{\star}}$ & 
$(T-T_r) \nabla \eta + (\mu-\mu_r) \nabla S$ \\  
\bottomrule
\end{tabularx}
\end{table}

\paragraph{\em Conventional Bernoulli function} 

Contrasting the ideal and compressible forms of the conventional Bernoulli function, 

\begin{equation}
      \frac{p + \rho g z}{\rho_{\star}}  \qquad \leftrightarrow \qquad 
      h(\eta,S,p) + g z  
      \label{conventional_bernoullis} 
\end{equation} 

\noindent 
makes it clear that the two expressions are not easily related. The difference in behaviour between $B_h$ and $B_h^{ideal}$ can be further evidenced by contrasting their vertical derivative for instance, leading to

\begin{equation}
    \frac{\partial}{\partial z} B_h^{ideal}  = 
   \frac{\partial}{\partial z} \left ( \frac{p+\rho g z}{\rho_\star} \right ) 
   = \frac{gz}{\rho_{\star}} \frac{\partial \rho}{\partial z} ,
\end{equation}

\begin{equation}
   \frac{\partial B_h}{\partial z} = 
   \frac{\partial}{\partial z} ( h + g z )  = 
   T \frac{\partial \eta}{\partial z} + \mu \frac{\partial S}{\partial z} .
   \label{dbhdz} 
\end{equation}

\noindent 
In order to link $\partial B_h/\partial z$ to the vertical derivative of some density variable one needs to make use of the Maxwell relations attached to the total differential of enthalpy ${\rm d}h = T {\rm d}\eta + \mu {\rm d}S + \upsilon {\rm d}p$, obtained by 
stating the equality of the cross derivatives, viz., 

\begin{equation}
     \frac{\partial^2 h}{\partial \eta \partial p} = 
     \frac{\partial T}{\partial p} = \frac{\partial \upsilon}{\partial \eta} 
     \qquad \frac{\partial^2 h}{\partial S \partial p} = 
     \frac{\partial \mu}{\partial p} = 
     \frac{\partial \upsilon}{\partial S} .
     \label{maxwell_relations} 
\end{equation}

\noindent 
To proceed, let us now introduce some reference pressure $p_{\star} = p_0(z_{\star})$ envisioned as being not too different from $p$, defining the reference values $T_{\star} = T(\eta,S,p_{\star}$ and $\mu_{\star} = \mu(\eta,S,p_{\star})$. If we also use use the approximation $p \approx p_0(z)$, (\ref{dbhdz}) may be rewritten as 

\begin{equation}
\begin{split} 
        \frac{\partial B_h}{\partial z} = & \int_{p_{\star}}^{p} 
        \frac{\partial \upsilon}{\partial \eta} \,{\rm d}p' 
        \frac{\partial \eta}{\partial z} 
        + \int_{p_{\star}}^p 
        \frac{\partial \upsilon}{\partial S} \,{\rm d}p'
        \frac{\partial S}{\partial z} 
        + T_{\star} \frac{\partial \eta}{\partial z} 
        + \mu_{\star} \frac{\partial S}{\partial z} \\
        \approx & (p-p_{\star} ) 
        \left ( \overline{\upsilon_{\eta}} \frac{\partial \eta}{\partial z}
         + \overline{\upsilon_S} \frac{\partial S}{\partial z} 
         \right ) + T_{\star} 
         \frac{\partial \eta}{\partial z}  + 
         \mu_{\star} \frac{\partial S}{\partial z} \\ 
         \approx & \frac{g(z-z_{\star})}{\rho_{\star}} 
         \frac{\partial \rho_{lr}}{\partial z} 
          + T_{\star} \frac{\partial \eta}{\partial z} 
          + \mu_{\star} \frac{\partial S}{\partial z} .
          \label{dbhdz_bis}
\end{split} 
\end{equation}

\noindent where $\rho_{lr}$ is the potential density referenced to $p_{\star}$ (or more accurately to $(p_{\star}+p)/2$), where the result assumed that $p_{\star}$ is close enough to $p$ that the overbar value can be approximated by their values at $p_{\star}$. While the first term of (\ref{dbhdz_bis}) succeeds in showing dependence on the vertical derivative of some density variable that makes it directly comparable to $dB_h^{ideal}/dz$, the second term is extraneous. Using the same ideas, it is also easily established that the same difficulties exist for relating ${\bf P}_h$ to ${\bf P}_h^{ideal}$. 

\paragraph{\em Available Bernoulli function} 

In contrast, it is immediately apparent that the ideal and compressible available Bernoulli functions 

\begin{equation}
      E_a +  \frac{p - p_0(z)}{\rho_{\star}}  \qquad 
       \leftrightarrow \qquad 
       \Pi_1 + \Pi_2 + \frac{p-p_0(z)}{\rho} 
       \label{available_bernoullis} 
\end{equation}

\noindent 
are directly comparable, as $B_a^{ideal}$ can be obtained from $B_a$ by neglecting $\Pi_1$ and replacing $\rho$ by $\rho_{\star}$ in $(p-p_0(z))/\rho$, as \citet{Tailleux2018} showed $\Pi_2$ to closely resemble $E_a$. Likewise, it is possible to show that the APE-based P-vectors resemble each other

\begin{equation}
     \frac{g (z-z_r) \nabla \rho}{\rho_{\star}} \qquad 
     \leftrightarrow \qquad  (T-T_r) \nabla \eta + (\mu-\mu_r) \nabla S 
\end{equation}

\noindent 
To see the connection, one may use the Maxwell relations (\ref{maxwell_relations}) to establish that 

\begin{equation}
\begin{split} 
    {\bf P}_a = & (T-T_r) \nabla \eta + (\mu-\mu_r) \nabla S  \\ 
    =  &   \int_{p_r}^p \frac{\partial \upsilon}{\partial \eta} 
    (\eta,S,p')\,{\rm d}p' \nabla \eta  + 
    \int_{p_r}^p \frac{\partial \upsilon}{\partial S} 
    (\eta,S,p')\,{\rm d}p' \nabla S  \\
    = & (p-p_r) \left ( \overline{\upsilon_{\eta}} \nabla \eta 
+ \overline{\upsilon_S} \nabla S \right ) \\
     \approx & \frac{g(z-z_r)}{\overline{\rho}} 
     \left ( \overline{\rho_{\eta}} \nabla \eta + 
     \overline{\rho_S} \nabla S \right ) 
\end{split}
\end{equation}

\noindent 
using the fact that $\upsilon_{\eta} = - \rho_{\eta}/\rho^2$, $\upsilon_S = -\rho_S/\rho^2$, and $p-p_r \approx -\overline{\rho} g (z-z_r)$, where $\upsilon_{\eta} = \partial \upsilon/\partial \eta$, $\upsilon_S = \partial \upsilon/\partial S$, etc...
This result shows that ${\bf P}_a$ is intermediate between the locally-referenced gradients referenced to the local and reference pressure respectively, which means that it is close to the standard neutral vector considered by \cite{McDougall1987}. In other words, ${\bf P}_a$ is approximately proportional to the gradient of the potential density referenced to the mid-pressure $(p+p_r)/2$. For more extensive discussion of this link, see \cite{Tailleux2023}.  

\par 
The above considerations clearly establish that the results of ideal fluid thermocline theory can only be generalised to compressible seawater if the available form $(B_a,{\bf P}_a)$ rather than the conventional form $(B_h,{\bf P}_h)$ is used.
 
\subsection{Inactive wind solutions}

To discuss how to extend Needler's formula to compressible seawater, we consider the steady and inviscid momentum equations written in their thermodynamic or Crocco-Vazsonyi form

\begin{equation}
      {\boldsymbol{\omega}}_a \times {\bf v} + \nabla B_{\ell} 
      = {\bf P}_{\ell} , 
      \label{steady_EGTBF}
\end{equation}

\noindent
(from Eqs. (\ref{crocco_NSE}) and (\ref{crocco_NSE_bis})),
where the subscript $\ell=a,h$ indicates whether the conventional or available form of $B$ and ${\bf P}$ is used. To simplify notations, this subscript is dropped in the following and re-introduced only when needed. For small Rossby number, as pertains to the large-scale motions of interest here, the relative vorticity and kinetic energy only affect $\boldsymbol{\omega}_a$ and $B$ at second order.
Eq. (\ref{steady_EGTBF}) is then an under-determined linear system for ${\bf v}$
that can only be solved if the following solvability condition is satisfied

\begin{equation}
    {\boldsymbol{\omega}}_a \cdot {\bf P} 
     = {\boldsymbol{\omega}}_a \cdot \nabla B ,
     \label{solvability_condition} 
\end{equation}

\noindent 
(obtained by taking the inner product of (\ref{steady_EGTBF}) with ${\boldsymbol{\omega}}_a$), so that 
${\bf v}$ in (\ref{steady_EGTBF}) is determined only up to an arbitrary vector $\chi \boldsymbol{\omega}_a$ parallel to $\boldsymbol{\omega}_a$, with $\chi$ a scalar field. The $\chi=0$ solution of (\ref{steady_EGTBF}) perpendicular to both $\nabla B$ and ${\bf P}$ and counterpart of Needler's formula (\ref{needler_formula}), is easily verified to be

\begin{equation}
       {\bf v} \stackrel{\text{def}}{=}  {\bf v}_{ia} = 
      \frac{{\bf P}\times
      \nabla B}{\rho Q_{b}} , 
       \label{full_solution} 
\end{equation}

\noindent 
while the counterpart of $Q=f\rho_z$ is

\begin{equation}
       Q_{b}  = \frac{{\boldsymbol{\omega}}_a \cdot \nabla B}{\rho} = \frac{\boldsymbol{\omega}_a \cdot {\bf P}}{\rho},
\end{equation}

\noindent 
the equality following from (\ref{solvability_condition}). In the context of a dry atmosphere, \cite{Gassmann2014} derived an expression similar to (\ref{full_solution}) and referred to ${\bf v}_{ia}$ as the {\em inactive wind}. 
\par 
In order to examine the consistency of the inactive wind solution (\ref{full_solution}) with mass conservation, let us take take the divergence of $\rho {\bf v}_{ia}$, which yields 

\begin{equation}
\begin{split} 
      \nabla \cdot (\rho {\bf v}_{ia} ) = &  
       - \frac{({\bf P} \times \nabla B)}{Q_b^2} 
       \cdot \nabla Q_b 
       + \frac{(\nabla \times {\bf P}) \cdot \nabla B}{Q_b} \\ 
     = & - \frac{\rho {\bf v}_{ia} \cdot \nabla Q_b}{Q_b} 
       + \frac{(\nabla \times {\bf P})\cdot \nabla B}{Q_b} .
%       = &     - \frac{\boldsymbol{\omega}_a \cdot \nabla }
%{\tilde{\boldsymbol{\omega}}_a\cdot \nabla B}
 %   \left ( \frac{DB}{Dt} \right ) = 0 ,
       \end{split} 
       \label{mass_conservation_via} 
\end{equation} 

\noindent 
As for one of Needler's formula, it can be shown that ${\bf v}_{ia}$ satisfies the continuity equation as a consequence of Ertel's PV conservation theorem applied to the PV constructed from the Bernoulli function $Q=\boldsymbol{\omega}_a\cdot \nabla B/\rho$. To show this, let us recall that in its most general form, \cite{Ertel1942}'s theorem (see \cite{Schubert2004} for an English translation) establishes that for any scalar $\lambda$, the PV variable $Q_{\lambda} = \boldsymbol{\omega}_a \cdot \nabla \lambda/\rho$ can be shown to satisfy the conservation law

\begin{equation}
    \frac{DQ_{\lambda}}{Dt} = \frac{\boldsymbol{\omega}_a}{\rho} 
    \cdot \nabla \left ( \frac{D\lambda}{Dt} \right ) + 
    \frac{1}{\rho^3} \nabla \lambda \cdot (\nabla \rho \times \nabla p)
    + \frac{\nabla \lambda \cdot \nabla \times {\bf F}}{\rho} 
    \label{PV_lambda} 
\end{equation}

\noindent 
e.g., see Eq. (4.95) of \cite{Vallis2006}. To make the link with (\ref{mass_conservation_via}), it is important to understand the different equivalent forms that the baroclinic production term (the term proportional to $\nabla \rho \times \nabla p$ in (\ref{PV_lambda})) may assume. As seen previously, from the definitions of ${\bf P}_h$ and ${\bf P}_a$, we have the following equivalence relations

\begin{equation} 
  \frac{1}{\rho} \nabla p + \nabla \Phi  
  = \nabla (h+\phi) - {\bf P}_h = \nabla (h+\Phi-B_r ) - {\bf P}_a .
\end{equation} 

\noindent 
Taking the curl yields the following equivalent expressions

\begin{equation} 
   \frac{1}{\rho^2} \nabla \rho \times \nabla p = \frac{1}{\rho^2} {\bf N}\times \nabla p =
   \nabla \times {\bf P}_h =  \nabla \times {\bf P}_a . 
\end{equation} 

\noindent 
which are proportional to the baroclinic production term, 
where ${\bf N} = \rho_S \nabla S + \rho_{\theta} \nabla \theta 
= \rho_{\eta} \nabla \eta + \rho_S \nabla S$ is the so called N-neutral vector entering \cite{McDougall1987}'s definition of (approximately) neutral surfaces (ANS). Now, if we use $B=\lambda$, in the inviscid case ${\bf F}=0$ and invoke Bernoulli theorem, $DB/Dt = 0$, (\ref{PV_lambda}) predicts that

\begin{equation}
     \frac{DQ_b}{Dt} = {\bf v}_{ia} \cdot \nabla Q_b = \frac{1}{\rho} \nabla B 
     \cdot (\nabla \times {\bf P}) 
\end{equation} 

\noindent 
Comparing this with (\ref{mass_conservation_via}) shows that

\begin{equation}
    \nabla \cdot ( \rho {\bf v}_{ia} ) = 0 ,
\end{equation}

\noindent 
as expected, thus confirming that ${\bf v}_{ia}$ satisfies the continuity equation regardless of which form of $(B,{\bf P})$ is used. 

\subsection{Uniqueness of the inactive wind solution}

The two inactive wind solutions introduced above may be written explicitly as

\begin{equation}
     {\bf v}_{ia}^h = \frac{{\bf P}_h \times \nabla B_h}
     {\boldsymbol{\omega}_a \cdot \nabla B_h} , \qquad
     {\bf v}_{ia}^a = \frac{{\bf P}_a \times \nabla B_a}
     {\boldsymbol{\omega}_a \cdot \nabla B_a} .
\end{equation}

\noindent 
Since these two solutions both satisfy the continuity equation, a question that naturally arises is whether these may actually define the same velocity field despite being constructed from different fields. To examine this, let us recall that by definition ${\bf P}_a = {\bf P}_h - \nabla B_r$, while $B_a = B_h - B_r$. It follows that 

\begin{equation}
\begin{split} 
     {\bf P}_a \times \nabla B_a = & ({\bf P}_h - \nabla B_r) \times 
     \nabla (B_h - B_r) \\
     = & {\bf P}_h \times \nabla B_h 
     + (\nabla B_h - {\bf P}_h ) \times \nabla B_r \\
     = & {\bf P}_h \times \nabla B_h 
      +\left ( \nabla \frac{{\bf v}^2}{2} +
      \frac{1}{\rho} \nabla_h p \right ) \times \nabla B_r ,
\end{split} 
\end{equation} 

\noindent 
which suggests that ${\bf v}_{ia}^h$ and ${\bf v}_{ia}^a$ define two different vector fields. However, because they both represent an exact solution of (\ref{steady_EGTBF}), it follows that their difference must be proportional to the null-space solution $\boldsymbol{\omega}_a$. In other words, there must exist some scalar $\delta \lambda$ such that

\begin{equation}
       {\bf v}_{ia}^h - {\bf v}_{ia}^a = \delta \lambda \,{\boldsymbol{\omega}}_a .
       \label{velocity_difference} 
\end{equation}

\noindent 
For small Rossby number, ${\boldsymbol{\omega}}_a \approx f {\bf k}$ (where $f$ is the Coriolis parameter), which implies that ${\bf v}_{ia}^h$ and ${\bf v}_{ia}^a$ must primarily differ by their vertical velocity component. Physically, this is plausible because in this case the horizontal components of ${\bf v}_{ia}^h$ and ${\bf v}_{ia}^a$ must both be approximately geostrophic, i.e., ${\bf u}_{ia}^h \approx {\bf u}_{ia}^a \approx {\bf u}_g$, while their vertical components must satisfy $w_{ia}^h = {\bf u}_g\cdot {\bf S}_h$ and $w_{ia}^a = {\bf u}_g\cdot {\bf S}_a$ respectively, where ${\bf S}_h$ and ${\bf S}_a$ are the horizontal slope vectors defined by ${\bf P}_h$ and ${\bf P}_a$ respectively. As a result,
\begin{equation}
        w_{ia}^h - w_{ia}^a = {\bf u}_g\cdot ({\bf S}_h - {\bf S}_a) ,
\end{equation}
which confirms that the two vertical velocities $w_{ia}^h$ and $w_{ia}^a$ might differ if the slopes defined by the two different P-vectors differ, as is indeed generally the case. It will be of interest to ascertain whether the vertical velocity component of ${\bf v}_{ia}^a$ is a better predictor of the actual vertical velocity field than that of ${\bf v}_{ia}^h$, which we plan on investigating in a subsequent study.

\subsection{Reformulation in terms of quantities independent of pressure}

Let $\gamma^T = \gamma^T(S,\theta)$ denote an approximately neutral quasi-material density variable and $Q_{\gamma T} = \boldsymbol{\omega}_a\cdot \nabla \gamma^T/\rho$ the PV variable constructed from it. As discussed in \cite{Saenz2015,Tailleux2016b,Tailleux2021,Tailleux2023}, APE theory naturally comes with a generalised form of potential density $\rho^{LZ}(S,\theta) = \rho(S,\theta,p_r)$ that is naturally very accurately neutral outside the Southern Ocean (even more so than Jackett and McDougall \cite{Jackett1997} empirical neutral density variable $\gamma^n)$, while also being mathematically and physically well defined. This motivated \cite{Tailleux2016b,Tailleux2021} to define thermodynamic neutral density $\gamma^T$ as 

\begin{equation}
    \gamma^T = \rho(S,\theta,p_r) - f_n(p_r)
\end{equation}

\noindent 
where $f_n(p_r)$ is a polynomial pressure correction empirically fitted to make $\gamma^T$ looks as much like $\gamma^n$. Because it tends to be more accurately neutral than $\gamma^n$ outside the Southern Ocean, $\gamma^T$ is the variable that is currently the least affected by the thermobaric production term.
\par 
If one accepts $\gamma^T$ and $Q_{\gamma T}$ as the most sensible generalisation of the concepts of density and PV to compressible seawater, then one may proceed similarly as in \cite{Needler1985} and assume that if $B$, $\gamma^T$ and $Q_{\gamma T}$ are all approximately conserved along fluid parcel trajectories, a functional relationship should exist between $B$, $\gamma^T$ and $Q_{\gamma T}$, say $B = G(\gamma^T,Q_{\gamma T})$ as in ideal fluid thermocline theory, then the seawater counterpart of  Needler's second formula (\ref{needler_formula_2}) becomes  

\begin{equation}
     {\bf v}_{ia} = \frac{\partial G}{\partial \gamma^T}
     \frac{{\bf P} \times \nabla \gamma^T}{\rho Q_b}
     + \frac{\partial G}{\partial Q_{\gamma T}}
      \frac{{\bf P} \times \nabla Q_{\gamma T}}{\rho Q_b} . 
      \label{via_linked} 
\end{equation}

\noindent 
In comparison to Needler's formula, (\ref{via_linked}) possesses the extra and undesirable thermobaricity-induced term proportional to $
{\bf P}\times \nabla \gamma^T$. This term can only be neglected if ${\bf P}_a$ rather than ${\bf P}_h$ is used, however, as the angle between ${\bf P}_h$ and $\nabla \gamma^T$ is not generally small enough. If so,

\begin{equation}
     {\bf v}_{ia} \approx \frac{\partial G}{\partial Q_{\gamma T}} \frac{{\bf P}_a \times \nabla Q_{\gamma T}}{\rho Q_b} ,
     \label{via_density_pv} 
\end{equation}

\noindent 
which is more directly comparable to (\ref{needler_formula_2}), stressing again the fundamental importance of APE theory to extend the results of ideal fluid thermocline theory to compressible seawater.

\section{Discussion} 
\label{discussion} 

The idea that steady fluid parcel trajectories lie at the intersection of the iso-surfaces of conserved quantities is arguably one of the most promising avenues of research for progressing the theory of the 3D oceanic velocity field, as it is one that is a priori as equally valid for an ideal fluid, for which it was originally developed by  \cite{Needler1985}, as well as for a fully compressible ocean with a realistic nonlinear equation of state, as recently initiated by \cite{Ochoa2020}. In this paper, we made significant progress towards generalising this idea to compressible seawater encapsulated into three main new results. 

\par 
Our first main result is that Needler's formula (\ref{needler_formula}) can be interpreted as a linear approximation of a much more general nonlinear and exact solution of the compressible NSE, called the 'inactive wind' solution. Such a solution was previously derived by \cite{Gassmann2014} in the context of a dry atmosphere and extended here to two-component compressible seawater. Like Needler's formula, the inactive wind solution satisfies the continuity equation as a consequence of Ertel's PV conservation theorem \cite{Muller1995} (that is itself related to the generalised Bernoulli theorem of \cite{Schar1993}). Like Needler's formula, the inactive wind solution is perpendicular to the gradient of the Bernoulli function, but unlike Needler's formula, it is perpendicular to a vector ${\bf P}$ rather than to the gradient of density $\nabla \rho$. Physically, the inactive wind solution is most easily obtained by rewriting the momentum equations in their Crocco-Vazsonyi or thermodynamic form, which appears to be the most illuminating form for relating the 3D velocity field to the conserved quantities of the fluid.

\par 
Our second main result is that both Needler's formula and the inactive wind solution are sensitive to how the Bernoulli function is defined, as it is always possible to redefine the latter by subtracting an arbitrary quasi-material function of $\rho$ (for a simple fluid) or of $S$ and $\theta$ (for compressible seawater). We find that only if the `available' form of Bernoulli function is used it is possible to meaningfully relate the inactive wind solution to Needler's formula. Physically, the available Bernoulli function is defined as the difference between the conventional Bernoulli function and its background reference value in Lorenz state of minimum potential energy entering Lorenz APE theory \cite{Lorenz1955}. Indeed, only in this case is the vector ${\bf P}$ parallel to an approximately neutral density variable, namely the Lorenz reference density (LRD) discussed at length in \cite{Tailleux2016b,Tailleux2021,Tailleux2023}. 

\par 

Our third main result is that inactive wind solutions defined for different forms of Bernoulli function and vector ${\bf P}$ do not necessarily define the same 3D velocity field even if each represents an exact solution of the compressible NSE satisfying the continuity equation. Mathematically, this is because the nonlinear balance equation from which the inactive wind solution is determined is degenerate. For small Rossby number, this is equivalent to say that different inactive wind solution differ primarily in their vertical velocity component. More generally, this result means that while it is in principle possible to construct a 3D velocity field as $\lambda \nabla C_1 \times \nabla C_2$ in terms of any arbitrary conserved quantities $C_1$ and $C_2$ for some $\lambda$, this does not necessarily imply that all constructions define the same velocity field, which does not appear to have received much attention so far. Physically, this is important because it provides the means, at least in principle, to test the usefulness of different constructions by comparing the 3D velocity field that each expression predicts against the 3D velocity field from any dynamically consistent ocean state estimate, as we plan on pursuing in a subsequent study. 

\par 

The present results are important because we believe that they can pave the way towards a more rigorous and general theory of the oceanic 3D velocity field valid for a realist compressible ocean, as we hope to further demonstrate through concrete applications in future studies.

%\begin{listing}[H]
%\caption{Title of the listing}
%\rule{\columnwidth}{1pt}
%\raggedright Text of the listing. In font size footnotesize, small, or normalsize. Preferred format: left aligned and single spaced. Preferred border format: top border line and bottom border line.
%\rule{\columnwidth}{1pt}
%\end{listing}

% Example of a page in landscape format (with table and table footnote).
%\startlandscape
%\begin{table}[H] %% Table in wide page
%\caption{This is a very wide table.\label{tab3}}
%	\begin{tabularx}{\textwidth}{CCCC}
%		\toprule
%		\textbf{Title 1}	& \textbf{Title 2}	& \textbf{Title 3}	& \textbf{Title 4}\\
%		\midrule
%		Entry 1		& Data			& Data			& This cell has some longer content that runs over two lines.\\
%		Entry 2		& Data			& Data			& Data\textsuperscript{1}\\
%		\bottomrule
%	\end{tabularx}
%	\begin{adjustwidth}{+\extralength}{0cm}
%		\noindent\footnotesize{\textsuperscript{1} This is a table footnote.}
%	\end{adjustwidth}
%\end{table}
%\finishlandscape

%%%%%%%%%%%%%%%%%%%%%%%%%%%%%%%%%%%%%%%%%%
\vspace{6pt} 

%%%%%%%%%%%%%%%%%%%%%%%%%%%%%%%%%%%%%%%%%%
%% optional
%\supplementary{The following supporting information can be downloaded at:  \linksupplementary{s1}, Figure S1: title; Table S1: title; Video S1: title.}


%%%%%%%%%%%%%%%%%%%%%%%%%%%%%%%%%%%%%%%%%%

\funding{This research has been supported by the NERC-funded OUTCROP project (grant no. NE/R010536/1).}

\dataavailability{The WOCE Global Ocean Climatology 1990-1998 (file '{\tt wghc\_params.nc}') used in this study is available at doi:10.25592/uhhfdm.8987. Enthalpy and other functions of state were estimated using the TEOS-10 library available at www.teos-10.org.  
Software to compute the analytic form of thermodynamic neutral density is available at
doi:10.5281/zenodo.4957697. See also https://github.com/GammaTN for regular software updates and illustrative code examples.} 

\acknowledgments{The author gratefully acknowledges comments from A. Colin de Verdi\`ere and Geoff Stanley on a original version of the paper, as well as constructive and supportive comments from three anonymous referees.}

\conflictsofinterest{The author declares no conflict of interest} 

%%%%%%%%%%%%%%%%%%%%%%%%%%%%%%%%%%%%%%%%%%
%% Optional

%% Only for journal Encyclopedia
%\entrylink{The Link to this entry published on the encyclopedia platform.}

\abbreviations{Abbreviations}{
The following abbreviations are used in this manuscript:\\

\noindent 
\begin{tabular}{@{}ll}
PV & Potential vorticity \\
APE & Available Potential Energy\\
ACE & Available Compressible Energy \\
LRD & Lorenz Reference Density \\ 
ANS & Approximately Neutral Surface \\
NSE & Navier-Stokes Equations 
\end{tabular}
}

%%%%%%%%%%%%%%%%%%%%%%%%%%%%%%%%%%%%%%%%%%
%% Optional
\appendixtitles{no} % Leave argument "no" if all appendix headings stay EMPTY (then no dot is printed after "Appendix A"). If the appendix sections contain a heading then change the argument to "yes".
\appendixstart
\appendix

%\section[Bernoulli theorems~\thesection]{Bernoulli theorem}

%\label{bernoulli_theorem_appendix} 

%It may be verified that the Bernoulli theorems satisfied by $B_a$ and %$B_h$ are: \\
%{\em Available Bernoulli theorem} 
%\begin{equation} 
 %   \frac{\partial B_a}{\partial t}
 %   +  {\bf v}\cdot \nabla B_a =
 %   \dotB_a + \frac{1}{\rho} 
  %  \frac{\partial \delta p}{\partial t} 
%\end{equation} 
%\begin{equation}
%     \dotB_a = (T-T_r) \dot{\eta} + 
 %    (\mu-\mu_r) \dot{S} + {\bf F} \cdot {\bf v} 
%\end{equation}
%{\em Standard Bernoulli theorem}
%\begin{equation}
%     \frac{\partial B_h}{\partial t}
%     + {\bf v}\cdot \nabla B_h 
%      =\dotB_h + \frac{1}{\rho} 
%      \frac{\partial p}{\partial t} 
%\end{equation}
%\begin{equation}
%       \dotB_h = T \dot{\eta} + \mu \dot{S}
%       + {\bf F}\cdot {\bf v} 
%\end{equation}

%%%%%%%%%%%%%%%%%%%%%%%%%%%%%%%%%%%%%%%%%%
\begin{adjustwidth}{-\extralength}{0cm}

\reftitle{References}


\bibliography{energetics}

\PublishersNote{}
\end{adjustwidth}
\end{document}